\documentclass{article}
\usepackage{spconf}

\usepackage{tikz}
\usepackage{pgfplots}
\usepackage{bm}
\usepackage{amsmath} 
\usepackage{amsthm}
\usepackage{amsfonts}
\usepackage{graphicx}
\usepackage{siunitx}
\usepackage{color}
\usepackage{fancyhdr}
\usepackage{array,multirow}
\usepackage{adjustbox}
\usepackage{balance}
\usepackage{cite}
\usepackage[acronym,shortcuts]{glossaries}
\usepackage{xspace}
\usepackage{tumcolor}
\usepackage{hyperref}
\usetikzlibrary{calc, shapes.geometric}

\fancypagestyle{cfooter}{ %
\fancyhf{} 
\cfoot{\footnotesize{\copyright\ 2023 IEEE. Personal use of this material is permitted. Permission from IEEE must be obtained for all other uses, in any current or future media, including reprinting/republishing this material for advertising or promotional purposes, creating new collective works, for resale or redistribution to servers or lists, or reuse of any copyrighted component of this work in other works.}}

}

\pgfplotsset{
	discard if not/.style 2 args={
		x filter/.code={
			\edef\tempa{\thisrow{#1}}
			\edef\tempb{#2}
			\ifx\tempa\tempb
			\else
			\fi
		}
	}
}
\pgfplotsset{compat=1.15}
\definecolor{mylila}{RGB}{153,50,204} 
\definecolor{mygreen}{RGB}{176,191,26} 

\def\pltw{240pt}
\def\plth{140pt}
\def\plthsmall{100pt}
\def\markSize{1.8pt}
\def\lineWidth{1.0pt}

\tikzset{VAEgenie/.style={mark options=solid, color=TUMBeamerRed, line width=\lineWidth, mark=Mercedes star, mark size=\markSize, solid}}
\tikzset{VAEnoisy/.style={mark options={solid}, color=TUMBeamerOrange, line width=\lineWidth, mark=triangle, mark size=\markSize, solid}}
\tikzset{VAEreal/.style={mark options={solid}, color=mygreen, line width=\lineWidth, mark=pentagon, mark size=\markSize, solid}}
\tikzset{VAErealvar/.style={mark options={solid}, color=mylila, line width=\lineWidth, mark=Mercedes star flipped, mark size=\markSize, solid}}
\tikzset{VAErealfix/.style={mark options={solid}, color=TUMGray, line width=\lineWidth, mark=square, mark size=\markSize, solid}}
\tikzset{geniecov/.style={mark options={solid}, color=blue, line width=\lineWidth, mark=x, mark size=\markSize, dashed}}
\tikzset{globalcov/.style={mark options={solid}, color=TUMBeamerGreen, line width=\lineWidth, mark=|, mark size=\markSize, dashed}}
\tikzset{LS/.style={mark options={solid}, color=black, line width=\lineWidth, mark=Mercedes star flipped, mark size=\markSize, dashed}}
\tikzset{GMM/.style={mark options={solid}, color=brown, line width=\lineWidth, mark=o, mark size=\markSize, dashed}}
\tikzset{{GMM circ}/.style={mark options={solid}, color=mylila, line width=\lineWidth, mark=diamond, mark size=\markSize, dashed}}
\tikzset{{GMM kron}/.style={mark options={solid}, color=brown, line width=\lineWidth, mark=o, mark size=\markSize, dashed}}
\tikzset{{GMM bcirc}/.style={mark options={solid}, color=mylila, line width=\lineWidth, mark=diamond, mark size=\markSize, dashed}}
\tikzset{CNN/.style={mark options={solid}, color=TUMBeamerLightBlue, line width=\lineWidth, mark=10-pointed star, mark size=\markSize, dashed}}
\tikzset{AMP/.style={mark options={solid}, color=TUMBeamerDarkRed, line width=\lineWidth, mark=star, mark size=\markSize, dashed}}
\tikzset{{genie OMP}/.style={mark options={solid}, color=TUMBeamerDarkRed, line width=\lineWidth, mark=star, mark size=\markSize, dashed}}
\tikzset{{globalcovlin}/.style={mark options={solid}, color=brown, line width=\lineWidth, mark=o, mark size=\markSize, dashed}}
\tikzset{{linint}/.style={mark options={solid}, color=TUMBeamerLightBlue, line width=\lineWidth, mark=10-pointed star, mark size=\markSize, dashed}}

\newcommand{\legVAEnoisy}{VAE-noisy}

\newcommand{\legVAErealvar}{VAE-real (var $\ma$)}
\newcommand{\legVAErealfix}{VAE-real (fix $\ma$)}
\newcommand{\leggeniecov}{CME}
\newcommand{\legglobalcov}{global}

\newcommand{\leggenieomp}{genie-OMP}
\newcommand{\legglobalcovlin}{global LI}
\newcommand{\leglinint}{LI}

\newcommand{\quadriga}{QuaDRiGa\xspace}

\newcommand{\Nrx}{{N_\text{rx}}}
\newcommand{\NL}{{N_\text{L}}}
\newcommand{\cnd}{{\,|\,}}

\newcommand{\msig}{{\bm{\Sigma}}}
\newcommand{\vdel}{{\bm{\delta}}}
\newcommand{\vmu}{{\bm{\mu}}}
\newcommand{\vsig}{{\bm{\sigma}}}

\newcommand{\vtheta}{{\bm{\theta}}}
\newcommand{\vphi}{{\bm{\phi}}}

\newcommand{\va}{{\bm{a}}}

\newcommand{\vc}{{\bm{c}}}

\newcommand{\vh}{{\bm{h}}}

\newcommand{\vn}{{\bm{n}}}

\newcommand{\vy}{{\bm{y}}}
\newcommand{\vz}{{\bm{z}}}

\newcommand{\ma}{{\bm{A}}}

\newcommand{\mc}{{\bm{C}}}

\newcommand{\mf}{{\bm{F}}}

\newcommand{\mh}{{\bm{H}}}

\newcommand{\mq}{{\bm{Q}}}

\newcommand{\RR}{{\mathbb{R}}}
\newcommand{\CC}{{\mathbb{C}}}
\newcommand{\diff}{{\mathrm{d}}}
\newcommand{\jim}{{\mathrm{j}}}
\newcommand{\OO}{{\mathcal O}}

\DeclareMathOperator{\E}{E}
\DeclareMathOperator{\vect}{vec}
\DeclareMathOperator{\diag}{diag}
\DeclareMathOperator{\KL}{D_{KL}}
\DeclareMathOperator*{\tran}{^{\mkern-2mu{T}}}
\DeclareMathOperator*{\herm}{^{\mkern-2mu{H}}}

\newacronym{tdd}{TDD}{time division duplex}
\newacronym{fdd}{FDD}{frequency division duplex}
\newacronym{lmmse}{LMMSE}{linear minimum mean squared error}
\newacronym{mse}{MSE}{mean squared error}
\newacronym{mmse}{MMSE}{minimum mean squared error}
\newacronym{nmse}{NMSE}{normalized mean squared error}
\newacronym{mimo}{MIMO}{multiple-input multiple-output}
\newacronym{simo}{SIMO}{single-input multiple-output}
\newacronym{miso}{MISO}{multiple-input single-output}
\newacronym{siso}{SISO}{single-input single-output}
\newacronym{deep}{DL}{deep learning}
\newacronym{ofdm}{OFDM}{orthogonal frequency division multiplexing}
\newacronym{csi}{CSI}{channel state information}
\newacronym{ula}{ULA}{uniform linear array}
\newacronym{ura}{URA}{uniform rectangular array}
\newacronym{dft}{DFT}{discrete fourier transform}
\newacronym{bs}{BS}{base station}
\newacronym{mt}{MT}{mobile terminal}
\newacronym{ae}{AE}{autoencoder}
\newacronym{ml}{ML}{machine learning}
\newacronym{dl}{DL}{deep learning}
\newacronym{doa}{DoA}{direction of arrival}
\newacronym{dod}{DoD}{direction of departure}
\newacronym{kl}{KL}{Kullback-Leibler}
\newacronym{elbo}{ELBO}{evidence-lower bound}
\newacronym{iid}{i.i.d.}{independent and identically distributed}
\newacronym{fc}{FC}{fully connected}
\newacronym{nn}{NN}{neural network}
\newacronym{dnn}{DNN}{deep neural network}
\newacronym{cnn}{CNN}{convolutional neural network}
\newacronym{ls}{LS}{least squares}
\newacronym{snr}{SNR}{signal-to-noise ratio}
\newacronym{ce}{CE}{channel estimation}
\newacronym{ul}{UL}{uplink}
\newacronym{mpc}{MPC}{multipath component}
\newacronym{rt}{RT}{ray-tracing}
\newacronym{mmd}{MMD}{maximum mean discrepancy}
\newacronym{cdf}{CDF}{cumulative distribution function}
\newacronym{tpr}{TPR}{true positive rate}
\newacronym{ccm}{CCM}{channel covariance matrix}
\newacronym{cg}{CG}{conditionally Gaussian}
\newacronym{3gpp}{3GPP}{3rd Generation Partnership Project}
\newacronym{vae}{VAE}{variational autoencoder}
\newacronym{vi}{VI}{variational inference}
\newacronym{cc}{CC}{convolutional channel}
\newacronym{cl}{CL}{convolutional layer}
\newacronym{ll}{LL}{linear layer}
\newacronym{rl}{RL}{reshaping layer}
\newacronym{bn}{BN}{batch normalization}
\newacronym{cs}{CS}{compressed sensing}
\newacronym{amp}{AMP}{approximate message passing}
\newacronym{gmm}{GMM}{Gaussian mixture model}
\newacronym{nf}{NF}{normalizing flow}
\newacronym{gan}{GAN}{generative adversarial network}
\newacronym{vdm}{VDM}{variational diffusion model}
\newacronym{cme}{CME}{conditional mean estimator}
\newacronym{los}{LOS}{line of sight}
\newacronym{nlos}{NLOS}{non-line of sight}
\newacronym{omp}{OMP}{orthogonal matching pursuit}
\newacronym{li}{LI}{linear interpolation}
\newacronym{mb}{MB}{model-based}
\newacronym{ud}{UD}{underdetermined}
\newacronym{fd}{FD}{fully-determined}

\title{Channel Estimation in Underdetermined Systems Utilizing \\ Variational Autoencoders}
\name{Michael Baur\thanks{This work is funded by the Bavarian Ministry of Economic Affairs, Regional Development and Energy within the project 6G Future Lab Bavaria. The authors acknowledge the financial support by the Federal Ministry of Education and Research of Germany in the program of “Souverän. Digital. Vernetzt.”. Joint project 6G-life, project identification number: 16KISK002.}, Nurettin Turan, Benedikt Fesl, Wolfgang Utschick}
\address{TUM School of Computation, Information and Technology, Technical University of Munich, Germany\\
Email: \{mi.baur, nurettin.turan, benedikt.fesl, utschick\}@tum.de}
\begin{document}
%
\maketitle

\thispagestyle{cfooter}

\begin{abstract}
In this work, we propose to utilize a \ac{vae} for \ac{ce} in \ac{ud} systems. The basis of the method forms a recently proposed concept in which a \ac{vae} is trained on \ac{csi} data and used to parameterize an approximation to the \ac{mse}-optimal estimator. The contributions in this work extend the existing framework from \ac{fd} to \ac{ud} systems, which are of high practical relevance. Particularly noteworthy is the extension of the estimator variant, which does not require perfect \ac{csi} during its offline training phase. This is a significant advantage compared to most other \ac{dl}-based \ac{ce} methods, where perfect \ac{csi} during the training phase is a crucial prerequisite. Numerical simulations for hybrid and wideband systems demonstrate the excellent performance of the proposed methods compared to related estimators.
\end{abstract}
\section{Introduction}
\label{sec:intro}

\Ac{dl} has the potential to further enhance future wireless communications systems~\cite{Bjornson2019}. 
For \ac{dl}, a \ac{dnn} is trained on site-specific data to capture prior information about a particular radio propagation environment, which can be leveraged to solve wireless communications-related problems.
In various prior work, \ac{dl}-based methods demonstrate excellent performance in \ac{ce}~\cite{Ye2018,Soltani2019,Neumann2018,Mashhadi2021,Baur2023}. 
In this context, \ac{mb}-\ac{dl} is an intriguing paradigm for the design of algorithms as it allows combining \ac{mb} insights with data-based learning~\cite{Shlezinger2023}.
Increased interpretability and a relaxed requirement for training data are two vital aspects \ac{mb} methods offer for novel algorithm development.

The recently proposed \ac{vae}-based channel estimator is a \ac{mb}-\ac{dl} method for \ac{ce}~\cite{Baur2022,Baur2023}.
The basic concept is to train a \ac{vae} on \ac{csi} data stemming from a radio propagation environment to learn the underlying channel distribution for which the \ac{vae} turns out to be ideally suited.
Subsequently, the trained \ac{vae} provides input data dependent conditional first and second moments to parameterize an approximation to the \ac{mse}-optimal \ac{cme}.
In~\cite{Baur2023}, the authors investigate an \ac{ul} \ac{fd} \ac{mimo} system. 
\Ac{ud} systems are not covered in~\cite{Baur2023}, but are of high practical relevance.
Typical \ac{ud} instances are hybrid systems with fewer RF chains than antennas~\cite{Ardah2018,Alkhateeb2014,Mendez-Rial2016,Koller2022a}, and wideband systems with time-evolving and frequency-selective channels~\cite{Fesl2022}.
Such systems have in common that perfect \ac{csi} data necessary for the training of most \ac{dl}-based methods are not available in real-world applications, apart from elaborate and costly measurement campaigns.
In sharp contrast, the VAE-real variant from~\cite{Baur2023} circumvents this significant drawback of \ac{dl}-based methods. 
Its training is solely based on noisy observations from \ac{fd} systems. 
However, it is an open question how the \ac{vae}-based estimators can be extended for applicability in \ac{ud} systems.

\textit{Contributions:} We propose to utilize a \ac{vae} for \ac{ce} in \ac{ud} systems.
We extend the approach from~\cite{Baur2023} to handle noisy observations stemming from a wide observation matrix, a typical aspect of \ac{ud} systems.
Of particular noteworthiness is the extension of the VAE-real variant, as it requires adapting the training procedure to deal with the information loss caused by the wide observation matrix. 
The non-trivial extensions in this work make sure that the VAE-real variant preserves its central advantage compared to other \ac{dl}-based \ac{ce} algorithms, which is the ability to be trained solely on noisy observations gathered during regular operation at the \ac{bs}.
Simulation results demonstrate the proposed methods' excellent \ac{ce} performance compared to related estimators.

\section{System and Channel Models}
\label{sec:system}

We assume to receive noisy observations $\vy$ of the channel $\vh$ as a linear map over a wide observation matrix $\ma$, i.e.,
\begin{equation}
    \vy = \ma\vh + \vn, \qquad \ma\in\CC^{M \times N},
    \label{eq:system-A}
\end{equation}
with $M<N$, $\vh \sim p(\vh)$, and $\vn\sim\mathcal{N}_{\CC}(\bm 0, \msig)$.
The matrix $\ma$ accounts for the available pilot information.
The task in \ac{ce} is to estimate $\vh$ based on $\vy$.
Ideally, we would like to evaluate the \ac{cme} $\E[\vh\cnd\vy]$, which yields \ac{mse}-optimal estimates~\cite[Ch.~11]{Kay1993}.
In general, a direct computation of the \ac{cme} is intractable as it requires access to the unknown distribution $p(\vh)$ and the solution of an integral over $\vh$, cf.~\cite{Baur2023}.
Another obstacle to performing \ac{ce} for the system model in~\eqref{eq:system-A} is that the matrix $\ma$ is wide, which makes the system \ac{ud}. 

\vspace{-1mm}
\subsection{Hybrid System}
\label{subsec:hybrid-arc}

Let us consider the \ac{ul} of a \ac{simo} system, i.e., we have a single-antenna \ac{mt} and a multi-antenna \ac{bs}.
In a hybrid system, the \ac{bs} has fewer RF chains $N_r$ than receive antennas $\Nrx$, and we have $M=N_r$ and $N=\Nrx$, cf.~\eqref{eq:system-A}.
It is usual to assume that the matrix $\ma$ either represents an analog phase-shift network or switches~\cite{Ardah2018,Alkhateeb2014,Mendez-Rial2016,Koller2022a}.
Since we cover the latter partly with the selection matrix in the wideband system model, cf. Section~\ref{subsec:wideband}, we let $\ma$ represent a phase-shift network.
To this end, $\ma$ is realized as a phase-shift matrix with a constant modulus constraint on the matrix elements~\cite{Koller2022a}.
A sub-Gaussian matrix comprises these properties and has favorable attributes for sparse recovery problems~\cite{Foucart2013}. 
More precisely, every entry $A_{i,k},\,i=1,\ldots,M,\,k=1,\ldots,N,$ of $\ma$ fulfills $ A_{i,k} = \frac{1}{\sqrt{M}} \exp(\jim\,\varphi),\,\varphi\sim\mathcal{U}([0, 2\pi])$.

We assume to have a \ac{ula} at the \ac{bs}, which enforces a Toeplitz-structured \ac{ccm}.
For the hybrid system, we make use of \acp{ccm} $\mc_\vdel$ that are defined according to a \ac{3gpp} spatial channel model for an urban macrocell scenario~\cite{3GPP2020}.
The spatial channel model computes the \ac{ccm} as $\mc_{\vdel}=\int_{-\pi}^{\pi} g(\vartheta;\vdel) \va(\vartheta) \va(\vartheta)\herm \diff \vartheta,$ where $g(\vartheta;\vdel)$ is an angular power spectrum and $\va(\vartheta)$ is the array steering vector of a \ac{ula}.
The variable $\vdel$ comprises information about the path gains and angles of arrival at the \ac{bs} and follows a prior distribution $p(\vdel)$, cf.~\cite{Neumann2018}.
We set the propagation cluster number to one in this work.
Once the \ac{ccm} $\mc_\vdel$ is computed, a channel sample is obtained as $\vh \mid \vdel \sim \mathcal{N}_{\CC}(\bm{0}, \mc_{\vdel})$.

\vspace{-1mm}
\subsection{Wideband System}
\label{subsec:wideband}

The time-evolution of a frequency-selective \ac{siso} channel is considered in the wideband system, where $\mh\in\CC^{N_c \times N_t}$ represents the time-frequency response of the channel with $N_c$ subcarriers and $N_t$ time slots.
Let $N_p$ be the number of pilots. 
The observation matrix is a selection matrix $\ma\in\{0,1\}^{N_p\times N_c N_t}$ with a single one in every row and $\vh=\vect(\mh)$.
The ones in $\ma$ represent the $N_p$ pilots at the corresponding positions of the $N_c N_t$ channel entries.
Thus, we have $M=N_p$ and $N=N_c N_t$, cf.~\eqref{eq:system-A}.

We generate wideband channels with the \quadriga channel simulator~\cite{Jaeckel2014}.
\quadriga computes the element in the $i$-th row and $j$-th column of the channel matrix $\mh$ as sum $H_{i,k} = \sum_{\ell=1}^L G_\ell \exp(-2\pi \jim f_i \tau_{k,\ell}),$ $i=1,\ldots,N_c,$ $k=1,\ldots,N_t$, over the paths $\ell$ of in total $L$ multi-paths.
The frequency of subcarrier $i$ is $f_i$, the delay of path $\ell$ at time step $k$ is $\tau_{k,\ell}$, and $G_\ell$ accounts for the path attenuation, antenna radiation pattern, and polarization.
We consider an urban macrocell scenario at 2.1\,GHz center frequency and 180\,kHz bandwidth with $N_c=12$ and $N_t=14$, a typical 5G wideband frame~\cite{3GPP2023}.
The \ac{bs} is at 25\,m height and covers a 120$^\circ$ sector. 
The \acp{mt} are between 35 and 500\,m away from the \ac{bs}, and are indoors (20\,\%) and outdoors (80\,\%).
An \ac{mt}'s velocity is uniformly distributed between 0 and 300\,km/h.

\section{VAE-based Channel Estimation}
\label{sec:vae_est}

\vspace{-1mm}
\subsection{Preliminaries about the VAE}
\label{subsec:prelim}

The elaborations in this section follow the principles of the \ac{vae}-based channel estimator in~\cite{Baur2023}.
The \ac{elbo} is the central term for the training of a \ac{vae} and is a lower bound to a parameterized likelihood model $p_\vtheta(\vh)$ of the unknown channel distribution $p(\vh)$.
An accessible version of the \ac{elbo} reads as~\cite{Kingma2019}
\begin{equation}
    \mathcal{L}_{\vtheta,\vphi}(\vh) = \E_{q_{\vphi}} \left[\log p_{\vtheta}(\vh\cnd\vz)\right] - \KL(q_{\vphi}(\vz\cnd\vy)\,\|\,p(\vz))
    \label{eq:vae}
\end{equation}
with $\E_{q_\vphi(\vz|\vy)}[\cdot] = \E_{q_\vphi}[\cdot]$ as the expectation according to the variational distribution $q_\vphi(\vz\cnd\vy)$, which depends on the latent vector $\vz\in\RR^\NL$, and is supposed to approximate $p_\vtheta(\vz\cnd\vh)$.
The last term in~\eqref{eq:vae} is the \ac{kl} divergence.

The \ac{vae} optimizes the \ac{elbo} with the help of \acp{dnn} and the reparameterization trick~\cite{Kingma2014}. 
To this end, it is necessary to define the involved distributions, which we fulfill as follows: $p(\vz) = \mathcal{N}(\bm{0},\mathbf{I}),$ $p_{\vtheta}(\vh\cnd\vz) = \mathcal{N}_{\CC}(\vmu_\vtheta(\vz),\mc_\vtheta(\vz)),$ and $q_{\vphi}(\vz\cnd\vy) = \mathcal{N}(\vmu_\vphi(\vy),\diag(\vsig^2_\vphi(\vy))).$
Both $p_{\vtheta}(\vh\cnd\vz)$ and $q_{\vphi}(\vz\cnd\vy)$ are therefore \ac{cg} distributions, realized by \acp{dnn} with parameters $\vtheta$ and $\vphi$.
Accordingly, we obtain closed-form expressions for the terms in~\eqref{eq:vae}, i.e., $\left( -\E_{q_{\vphi}} \left[\log p_{\vtheta}(\vh\cnd\vz)\right] \right)$ is replaced by the estimate
\begin{equation}
     \log\det(\pi\,\mc_\vtheta(\tilde\vz)) + (\vh - \vmu_\vtheta(\tilde\vz))\herm \mc^{-1}_\vtheta(\tilde\vz) (\vh - \vmu_\vtheta(\tilde\vz)),
    \label{eq:vae-dec-like}
\end{equation}
$\tilde\vz\sim q_{\vphi}(\vz\cnd\vy)$ and $\KL(q_\vphi(\vz\cnd\vy)\,\|\,p_\vtheta(\vz))$ is the \ac{kl} divergence between two Gaussian distributions, which has a closed-form expression~\cite{Baur2023}.
A \ac{vae} consists of an encoder that represents $q_\vphi(\vz\cnd\vy)$ and outputs $\{\vmu_\vphi(\vy),\vsig_\vphi(\vy)\}$ and a decoder that represents $p_{\vtheta}(\vh\cnd\vz)$ and outputs $\{\vmu_\vtheta(\vz),\mc_\vtheta(\vz)\}$.

\vspace{-1mm}
\subsection{Fully-Determined Case}
\label{subsec:fully-determined}

The \ac{vae}'s goal is to compose a \ac{cg} channel via the latent vector $\bm{z}$ such that $\vh \cnd   \vz \sim p_\vtheta(\vh\cnd\vz).$
As outlined in~\cite{Baur2023}, the law of total expectation allows us to reformulate the \ac{cme} as
\begin{equation}
    \E[\vh\cnd\vy] = \E_{\vz}[\E[\vh\cnd\vz,\vy]\cnd\vy].
    \label{eq:total_exp}
\end{equation}
Under the \ac{vae} framework, the inner expectation in~\eqref{eq:total_exp} exhibits a closed-form solution because the \ac{vae} models $\vh$ as \ac{cg}. 
The outer expectation is approximated by sampling from $q_{\vphi}(\vz\cnd\vy)$.
Let ${t_\vtheta(\vz,\vy) = \E[\vh\cnd\vz,\vy]}$, then $t_\vtheta(\vz,\vy) = $
\begin{equation}
    \vmu_\vtheta(\vz) + \mc_\vtheta(\vz) \ma\herm (\ma\mc_\vtheta(\vz)\ma\herm + \msig)^{-1} (\vy - \ma\vmu_\vtheta(\vz)).
    \label{eq:lmmse-vae}
\end{equation}
If we take one sample $\vz^{(1)}=\vmu_\vphi(\vy)$ to approximate the outer expectation in~\eqref{eq:total_exp}, we obtain the \ac{vae}-based channel estimator from~\cite{Baur2023}: $\hat\vh_{\text{VAE}}(\vy) = t_\vtheta( \vz^{(1)}=\vmu_\vphi(\vy), \vy)$.
Taking only the single sample $\vmu_\vphi(\vy)$ to compute $\hat\vh_{\text{VAE}}(\vy)$ achieves excellent \ac{ce} results for \ac{fd} systems~\cite{Baur2023}.
The estimator $\hat\vh_{\text{VAE}}(\vy)$ is non-linear as a result of the non-linear transformations by the encoder and decoder \acp{dnn}, providing $\vmu_\vtheta(\vz)$ and $\mc_\vtheta(\vz)$ depending on $\vy$.
It is proved in~\cite{Baur2023} that a \ac{vae}-based estimator is theoretically capable of converging to the true \ac{cme} under certain assumptions. 
One of the assumptions is that $\ma$ is invertible, meaning that the proof does not hold for a wide $\ma$, which motivates the investigation of \ac{ud} systems.

Among the practicable variants of \ac{vae}-based estimators are the VAE-noisy and VAE-real variants~\cite{Baur2023}.
For VAE-noisy, $\vy$ is the encoder input, and perfect \ac{csi} data $\vh$ is available during the training for the computation of~\eqref{eq:vae-dec-like}.
VAE-real has only access to $\vy$ during the evaluation \textit{and} training.
Hence, the \ac{elbo} is a lower bound to $p_\vtheta(\vy)$ and $p_{\vtheta}(\vh\cnd\vz)$ in~\eqref{eq:vae} becomes $p_{\vtheta}(\vy\cnd\vz)$ for VAE-real.
Consequently, this exchanges $\vh$ with $\vy$, $\vmu_\vtheta(\vz)$ with $\ma\vmu_\vtheta(\vz)$, and $\mc_\vtheta(\vz)$ with $\ma\mc_\vtheta(\vz)\ma\herm + \msig$ in the expression in~\eqref{eq:vae-dec-like}.
Since the decoder outputs $\mc_\vtheta(\vz)$, the VAE-real variant is still suitable for \ac{ce}.
An intriguing aspect concerning VAE-real is that its training is possible solely based on noisy observations $\vy$, in contrast to the plethora of \ac{dl}-based channel estimators that require access to perfect \ac{csi} data during their training phase.
VAE-real can, therefore, be trained at the BS side during regular \ac{bs} operation, making it a highly realistic estimator.

\vspace{-1mm}
\subsection{Underdetermined Case}
\label{subsec:underdetermined}

In general, if $\ma$ is wide, then the linear map $\ma\vh$ causes a loss of information about $\vh$ since a subset of the channels lies in the nullspace of $\ma$. 
For smaller $M$, more information is lost, which must be recovered during the training of a potential \ac{vae}-based channel estimator.
For the VAE-noisy variant, the loss computation is as in~\eqref{eq:vae-dec-like}, since it is assumed that $\vh$ is available during the training in this case.
Possible simplifications of the loss apply because of the parameterization of $\mc_\vtheta(\vz)$, discussed at the end of this section.
As encoder input for VAE-noisy, we select $\ma\herm\vy$ to obtain an input of the same size as $\vh$.
This encoder input does not contain the full information about $\vh$ due to the multiplication with $\ma$.
Fortunately, the entire information about the ground truth $\vh$ is still provided in the loss, which makes it possible that the \ac{vae} is capable of inferring a suitable latent representation for $\vh$ based on $\ma\herm\vy$ for the estimation task after the training.

The VAE-real variant uses the same encoder input as VAE-noisy.
The loss computation is different since only $\vy$ is available during the training. 
The non-invertibility of $\ma$ requires us to calculate~\eqref{eq:vae-dec-like} directly during the training as the multiplication with $\ma$ prevents possible simplifications enabled by $\mc_\vtheta(\vz)$.
However, all the additional computational complexity only affects the offline training procedure.
To compensate for the loss of information caused by $\ma$, we propose introducing a changing matrix $\ma$ during the training of VAE-real.
This way, the \ac{vae} can observe the full channel space over the training iterations. 
For the hybrid system, we do this by sampling a new $\ma$ in each iteration.
For the wideband system, we place the pilots randomly.
After the successful training, VAE-noisy and VAE-real each provide $\vmu_\vtheta(\vz)$ and $\mc_\vtheta(\vz)$ to determine the estimate $\hat\vh_{\text{VAE}}(\vy)$.
What remains is to clarify how $\mc_\vtheta(\vz)$ is parameterized for the considered system models.

\textit{Hybrid System:} Since we assume to have a \ac{ula} at the \ac{bs} in the hybrid system, the true \ac{ccm} is a Toeplitz matrix.
A Toeplitz \ac{ccm} can be asymptotically approximated by a circulant matrix for a large number of antennas~\cite{Gray2005}, which is present in massive \ac{mimo} systems.
Accordingly, we model the \ac{ccm} $\mc_\vtheta(\vz)$ as a circulant matrix.
Furthermore, a circulant \ac{ccm} is diagonalizable by the \ac{dft} matrix $\mf\in\CC^{N \times N}$ such that $\mc_\vtheta(\vz) = \mf\herm\diag(\vc_\vtheta(\vz))\mf,\, \vc\in\RR_+^N,$
which implies that we can invert $\mc_\vtheta(\vz)$ in $\OO(N\log N)$ time.
As a result, we only need to learn a positive and real-valued vector $\vc_\vtheta(\vz)$ at the \ac{vae} decoder to parameterize a full covariance matrix.

\textit{Wideband System:} Assuming a constant time-sampling and carrier-spacing, the \acp{ccm} along the time- and frequency-axis are each reasonably assumed to be Toeplitz-structured. 
As $N_t=14$ and $N_c=12$, a circulant approximation to each of the \acp{ccm} is unfavorable because of the relatively short time and frequency windows.
Hence, we model $\mc_\vtheta(\vz)$ as a block-Toeplitz matrix $\mc_\vtheta(\vz) = \mc_{\vtheta,t}(\vz) \otimes \mc_{\vtheta,c}(\vz) = \mq\herm \diag(\vc_\vtheta(\vz)) \mq,$
$\vc_\vtheta(\vz)\in\RR_+^{4N_c N_t},$ where $\mc_{\vtheta,t}(\vz)$ represents the Toeplitz \ac{ccm} along the time axis, and $\mc_{\vtheta,c}(\vz)$ along the frequency axis. \
With $\mq = \mq_t \otimes \mq_c$, the matrix $\mq_t\in\CC^{2N_t\times N_t}$ contains the first $N_t$ columns of the $2N_t \times 2N_t$ \ac{dft} matrix, and $\mq_c$ is defined analogously.
The matrix $\mq$, together with $\vc_\vtheta(\vz)$, defines a positive-definite Hermitian Toeplitz matrix and therefore motivates our parameterization choice for $\mc_\vtheta(\vz)$ for the wideband system~\cite{Fesl2022}.

\section{Simulation Results}
\label{sec:sim}

We create a training, validation, and test channel dataset of $T_{\text{r}}=180{,}000$, $T_{\text{v}}=10{,}000$, and $T_{\text{e}}=10{,}000$ samples.
We assume that $\msig=\varsigma^2\mathbf{I}$ and knowledge of the noise variance $\varsigma^2$ during the training and estimation.
The channels are normalized in each dataset such that $\E[\|\vh\|^2]=N$. 
Accordingly, we define the \ac{snr} as $1/\varsigma^2$.
We evaluate our methods with the \ac{nmse} $\frac{1}{T_{\text{e}}N} \sum_{i=1}^{T_{\text{e}}}\|\vh_i-\hat{\vh}_i\|^2$, where $\vh_i$ is the $i$-th test channel realization, and $\hat{\vh}_i$ the corresponding estimate.
The \ac{vae} architectures are almost equivalent to the ones in~\cite{Baur2023}, meaning we use 1D convolutional layers, batch normalization, ReLU activations, and linear layers to map between dimensions.
For a more detailed description, we refer the reader to~\cite{Baur2023} and the available simulation code\footnote{\url{https://github.com/baurmichael/vae-est-ud}.}.
The hyperparameters for the architectures were obtained with a random search~\cite{Bergstra2012}.

We consider the following baseline channel estimators for the hybrid system.
As we know the \ac{ccm} $\mc_\vdel$ in the \ac{3gpp} channel model, we can evaluate the \ac{cme}: $\hat\vh_{\text{CME}}(\vy) = \mc_\vdel \ma\herm (\ma\mc_\vdel\ma\herm + \msig)^{-1} \vy$. 
We can also compute a global sample covariance matrix $\hat\mc=\frac{1}{T_\text{r}}\sum_{i=1}^{T_\text{r}} \vh_i\vh_i^{\text H}$ from the training dataset channels to evaluate $\hat\vh_{\text{global}}(\vy) = \hat\mc \ma\herm (\ma\hat\mc\ma\herm + \msig)^{-1} \vy$.
Lastly, we also compute an estimate with the \ac{omp} algorithm, a prominent representative from the \ac{cs} literature~\cite{Alkhateeb2014,Mendez-Rial2016,Koller2022a}.
We utilize a two times oversampled \ac{dft} matrix as a dictionary for \ac{omp} as a result of the \ac{ula} at the \ac{bs} and use the true channel to determine the sparsity order, which is why we call this estimator genie-OMP.

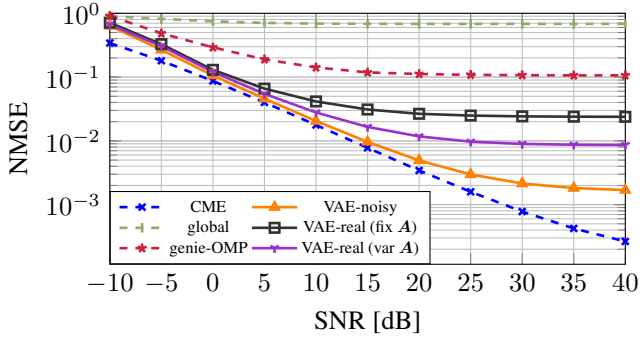
\begin{figure}
    \centering
    \begin{tikzpicture}
    	\begin{semilogyaxis}[
    		xlabel={SNR [dB]},
    		ylabel={NMSE},
    		xmin=-10, xmax=40,
            xtick={-10,-5,0,5,10,15,20,25,30,35,40},
    		ymajorgrids=true,
    		xmajorgrids=true,
    		grid=both, 
    		grid style=solid,
    		legend pos=south west,
    		width=\pltw,
    		height=\plth,
    		legend style={nodes={scale=0.63, transform shape}},
    		legend columns=2,
    		ymax=1,
		    legend style={at={(0,0)},anchor=south west}
    	] 
    	\addplot[geniecov]
    		table [ignore chars=", x=snr, y=mse, col sep=comma] {data/results-mse-3gpp-hybrid-learn-1p-128rx-32rf-genie_cov.txt};
    	\addlegendentry{\leggeniecov}
    	
    	\addplot[VAEnoisy]
    		table [ignore chars=", x=snr, y=mse, col sep=comma] {data/results-mse-3gpp-hybrid-learn-1p-128rx-32rf-vae_noisy.txt};
    	\addlegendentry{\legVAEnoisy}
    	
    	\addplot[globalcov]
    		table [ignore chars=", x=snr, y=mse, col sep=comma] {data/results-mse-3gpp-hybrid-learn-1p-128rx-32rf-global_cov.txt};
    	\addlegendentry{\legglobalcov}
    	
    	\addplot[VAErealfix]
    		table [ignore chars=", x=snr, y=mse, col sep=comma] {data/results-mse-3gpp-hybrid-learn-1p-128rx-32rf-vae_real_fix.txt};
    	\addlegendentry{\legVAErealfix}
    	
    	\addplot[genie OMP]
    		table [ignore chars=", x=snr, y=mse, col sep=comma] {data/results-mse-3gpp-hybrid-learn-1p-128rx-32rf-genie_omp.txt};
    	\addlegendentry{\leggenieomp}
    	
    	\addplot[VAErealvar]
    		table [ignore chars=", x=snr, y=mse, col sep=comma] {data/results-mse-3gpp-hybrid-learn-1p-128rx-32rf-vae_real_vary.txt};
    	\addlegendentry{\legVAErealvar}
    	
    	\end{semilogyaxis}
\end{tikzpicture}
    \vspace{-6mm}
    \caption{\ac{nmse} over the \ac{snr} for the hybrid system from Section~\ref{subsec:hybrid-arc} with $N=128$ antennas and $N_r=32$ RF chains.}
    \label{fig:hybrid_nmse}
    \vspace{-3mm}
\end{figure}

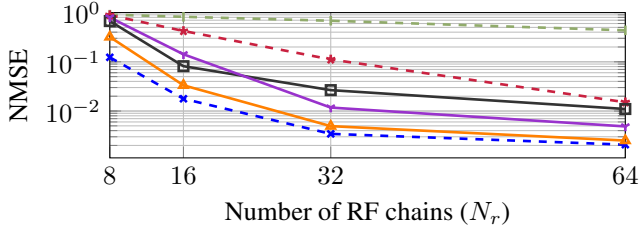
\begin{figure}
    \centering
    \begin{tikzpicture}
    	\begin{semilogyaxis}[
    		xlabel={Number of RF chains ($N_r$)},
    		ylabel={NMSE},
    		xmin=8, xmax=64,
            xtick={8,16,32,64},
    		ymajorgrids=true,
    		xmajorgrids=true,
    		grid=both,
    		grid style=solid,
    		legend pos=south west,
    		width=\pltw,
    		height=\plthsmall,
    		legend style={nodes={scale=0.63, transform shape}},
    		legend columns=2,
    		ymax=1,
		    legend style={at={(0,0)},anchor=south west}
    	] 
    	\addplot[geniecov]
    		table [ignore chars=", x=rf, y=genie_cov, col sep=comma]{data/nmse_over_rf_20db.txt};
    	
    	\addplot[globalcov]
    		table [ignore chars=", x=rf, y=global_cov, col sep=comma] {data/nmse_over_rf_20db.txt};
    	
    	\addplot[genie OMP]
    		table [ignore chars=", x=rf, y=genie_omp, col sep=comma] {data/nmse_over_rf_20db.txt};
    	
    	\addplot[VAEnoisy]
    		table [ignore chars=", x=rf, y=vae_noisy, col sep=comma] {data/nmse_over_rf_20db.txt};
    	
    	\addplot[VAErealfix]
    		table [ignore chars=", x=rf, y=vae_real_fix, col sep=comma]{data/nmse_over_rf_20db.txt};
    	
    	\addplot[VAErealvar]
    		table [ignore chars=", x=rf, y=vae_real_var, col sep=comma]{data/nmse_over_rf_20db.txt};
    	
    	\end{semilogyaxis}
\end{tikzpicture}
    \vspace{-6mm}
    \caption{\ac{nmse} over the RF chains for the hybrid system from Section~\ref{subsec:hybrid-arc} with $N=128$ antennas at an \ac{snr} of 20\,dB.}
    \label{fig:hybrid_rf}
    \vspace{-3mm}
\end{figure}

In Fig.~\ref{fig:hybrid_nmse}, we display the \ac{nmse} over the \ac{snr} for the hybrid system from Section~\ref{subsec:hybrid-arc} with $N=128$ antennas and $N_r=32$ RF chains. 
It is visible that all estimators, except the CME, exhibit a saturation in the high \ac{snr}, which relates to the information loss caused by $\ma$. 
After the utopian CME, VAE-noisy performs best, followed by VAE-real with a varying $\ma$ during the training (purple curve).
Consequently, the variation of $\ma$ during the training of VAE-real improves the estimation performance.
Although genie-OMP uses utopian channel knowledge during the estimation, it performs poorly.
The $\hat\vh_\text{global}(\vy)$ estimator performs worst.
We illustrate the performance for a varying number of RF chains in Fig.~\ref{fig:hybrid_rf} with the same simulation parameters and methods as in Fig.~\ref{fig:hybrid_nmse}.
All methods improve in terms of \ac{nmse} for increasing $N_r$.
The qualitative order of the methods is as in Fig.~\ref{fig:hybrid_nmse}. 
Except for $N_r<32$, the VAE-real with varied $\ma$ during the training outperforms the VAE-real with a fixed $\ma$.

\begin{figure}
    \centering
    \begin{tikzpicture}
    	\begin{semilogyaxis}[
    		xlabel={SNR [dB]},
    		ylabel={NMSE},
    		xmin=-10, xmax=40,
    		ymax=1, ymin=1e-4,
    		ytick={1e-4,1e-3,1e-2,1e-1,1},
            xtick={-10,-5,0,5,10,15,20,25,30,35,40},
    		ymajorgrids=true,
    		xmajorgrids=true,
    		grid=both,
    		grid style=solid,
    		legend pos=south west,
    		width=\pltw,
    		height=\plth,
    		legend style={nodes={scale=0.63, transform shape}},
    		legend columns=2,
		    legend style={at={(0,0)},anchor=south west}
    	] 
    	\addplot[globalcov]
    		table [ignore chars=", x=snr, y=mse, col sep=comma] {data/results-mse-quadriga-wideband-mixed-12c-14t-20Np-global_cov.txt};
    	\addlegendentry{\legglobalcov}
    	
    	\addplot[VAEnoisy]
    		table [ignore chars=", x=snr, y=mse, col sep=comma] {data/results-mse-quadriga-wideband-mixed-12c-14t-20Np-vae_noisy.txt};
    	\addlegendentry{\legVAEnoisy}
    	
    	\addplot[globalcovlin]
    		table [ignore chars=", x=snr, y=mse, col sep=comma]{data/results-mse-quadriga-wideband-mixed-12c-14t-20Np-global_cov_lin.txt};
    	\addlegendentry{\legglobalcovlin}
    	
    	\addplot[VAErealfix]
    		table [ignore chars=", x=snr, y=mse, col sep=comma] {data/results-mse-quadriga-wideband-mixed-12c-14t-20Np-vae_real_fix.txt};
    	\addlegendentry{\legVAErealfix}
    	
    	\addplot[linint]
    		table [ignore chars=", x=snr, y=mse, col sep=comma] {data/results-mse-quadriga-wideband-mixed-12c-14t-20Np-lin_int.txt};
    	\addlegendentry{\leglinint}
    	
    	\addplot[VAErealvar]
    		table [ignore chars=", x=snr, y=mse, col sep=comma] {data/results-mse-quadriga-wideband-mixed-12c-14t-20Np-vae_real_vary.txt};
    	\addlegendentry{\legVAErealvar}
    	
    	\end{semilogyaxis}
\end{tikzpicture}
    \vspace{-6mm}
    \caption{\ac{nmse} over the \ac{snr} for the wideband system from Section~\ref{subsec:wideband} with $N_p=20$ pilots in lattice layout.}
    \label{fig:wideband_nmse}
    \vspace{-3mm}
\end{figure}
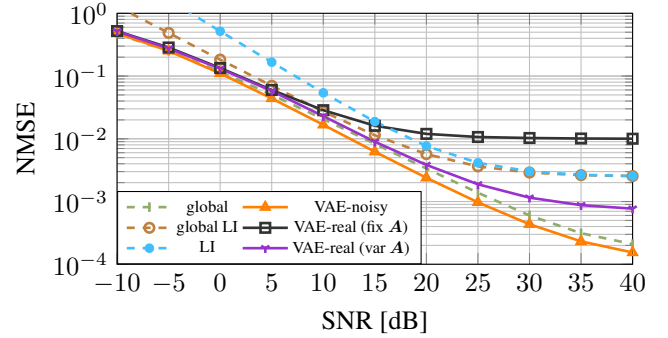

At last, in Fig.~\ref{fig:wideband_nmse}, we show the \ac{nmse} over the \ac{snr} for the wideband system from Section~\ref{subsec:wideband} with $N_p=20$ pilots in lattice layout as described in~\cite{Fesl2022}.
For a fair comparison with the VAE-real variants, we display the \ac{nmse} for an estimate obtained with \ac{li} of the missing entries in $\vy$ after computing $\ma\tran\vy$.
We also compute a sample covariance matrix for the channels from the \ac{li} estimates by subtracting the noise covariance from it and truncating the negative eigenvalues to subsequently perform a linear MMSE estimate (global LI) as for $\hat\vh_\text{global}(\vy)$.
All methods that work only with $\vy$ saturate in Fig.~\ref{fig:wideband_nmse} for high \ac{snr}.
VAE-noisy shows the best results for all \ac{snr} values, followed by the $\hat\vh_\text{global}(\vy)$ estimator.
VAE-real with varying $\ma$ during the training outperforms global \ac{li}, especially in the high \ac{snr}.
The performance gap between the two VAE-real variants for high \ac{snr} is more prominent in Fig.~\ref{fig:wideband_nmse} than in Fig.~\ref{fig:hybrid_nmse}.
It seems that varying the selection matrix $\ma$ during the training has a more beneficial impact, potentially due to its masking properties.

\vfill
\section{Conclusion}
\label{sec:conclusion}

This work proposes a \ac{vae}-based channel estimator for \ac{ud} systems.
To this end, we analyze hybrid and wideband systems, which are \ac{ud} systems of high relevance.
The \ac{vae}-based estimators exhibit superior \ac{ce} performance compared to the baseline estimators that have equivalent \ac{csi} knowledge for the design of the respective estimator.
The strong estimation results of the VAE-real variant are particularly noteworthy because VAE-real is trained solely with noisy observations gathered during regular \ac{bs} operation.
We plan to extend the evaluation to measurement data in our future work.

\clearpage
\balance\bibliographystyle{IEEEbib}
\bibliography{main}

\begin{thebibliography}{10}

\bibitem{Bjornson2019}
Emil Bj{\"{o}}rnson, Luca Sanguinetti, Henk Wymeersch, Jakob Hoydis, and
  Thomas~L. Marzetta,
\newblock ``{Massive MIMO is a reality—What is next?: Five promising research
  directions for antenna arrays},''
\newblock {\em Digit. Signal Process.}, vol. 94, pp. 3--20, 2019.

\bibitem{Ye2018}
Hao Ye, Geoffrey~Y. Li, and Biing-Hwang Juang,
\newblock ``{Power of Deep Learning for Channel Estimation and Signal Detection
  in OFDM Systems},''
\newblock {\em IEEE Wirel. Commun. Lett.}, vol. 7, no. 1, pp. 114--117, 2018.

\bibitem{Soltani2019}
Mehran Soltani, Vahid Pourahmadi, Ali Mirzaei, and Hamid Sheikhzadeh,
\newblock ``{Deep Learning-Based Channel Estimation},''
\newblock {\em IEEE Commun. Lett.}, vol. 23, no. 4, pp. 2019--2022, 2019.

\bibitem{Neumann2018}
David Neumann, Thomas Wiese, and Wolfgang Utschick,
\newblock ``{Learning The MMSE Channel Estimator},''
\newblock {\em IEEE Trans. Signal Process.}, vol. 66, no. 11, pp. 2905--2917,
  2018.

\bibitem{Mashhadi2021}
Mahdi~B. Mashhadi and Deniz Gunduz,
\newblock ``{Pruning the Pilots: Deep Learning-Based Pilot Design and Channel
  Estimation for MIMO-OFDM Systems},''
\newblock {\em IEEE Trans. Wirel. Commun.}, vol. 20, no. 10, pp. 6315--6328,
  2021.

\bibitem{Baur2023}
Michael Baur, Benedikt Fesl, and Wolfgang Utschick,
\newblock ``{Leveraging Variational Autoencoders for Parameterized MMSE Channel
  Estimation},''
\newblock {\em arXiv preprint arXiv:2307.05352}, 2023.

\bibitem{Shlezinger2023}
Nir Shlezinger, Jay Whang, Yonina~C. Eldar, and Alexandros~G. Dimakis,
\newblock ``{Model-Based Deep Learning},''
\newblock {\em Proc. IEEE}, vol. 111, no. 5, pp. 465--499, 2023.

\bibitem{Baur2022}
Michael Baur, Benedikt Fesl, Michael Koller, and Wolfgang Utschick,
\newblock ``{Variational Autoencoder Leveraged MMSE Channel Estimation},''
\newblock in {\em 2022 56th Asilomar Conf. Signals, Syst. Comput.}, Pacific
  Grove, CA, USA, 2022, pp. 527--532, IEEE.

\bibitem{Ardah2018}
Khaled Ardah, Gabor Fodor, Yuri C.~B. Silva, Walter~C. Freitas, and Francisco
  R.~P. Cavalcanti,
\newblock ``{A Unifying Design of Hybrid Beamforming Architectures Employing
  Phase Shifters or Switches},''
\newblock {\em IEEE Trans. Veh. Technol.}, vol. 67, no. 11, pp. 11243--11247,
  2018.

\bibitem{Alkhateeb2014}
Ahmed Alkhateeb, Omar {El Ayach}, Geert Leus, and Robert~W. Heath,
\newblock ``{Channel Estimation and Hybrid Precoding for Millimeter Wave
  Cellular Systems},''
\newblock {\em IEEE J. Sel. Top. Signal Process.}, vol. 8, no. 5, pp. 831--846,
  2014.

\bibitem{Mendez-Rial2016}
Roi Mendez-Rial, Cristian Rusu, Nuria Gonzalez-Prelcic, Ahmed Alkhateeb, and
  Robert~W. Heath,
\newblock ``{Hybrid MIMO Architectures for Millimeter Wave Communications:
  Phase Shifters or Switches?},''
\newblock {\em IEEE Access}, vol. 4, pp. 247--267, 2016.

\bibitem{Koller2022a}
Michael Koller and Wolfgang Utschick,
\newblock ``{Learning a compressive sensing matrix with structural constraints
  via maximum mean discrepancy optimization},''
\newblock {\em Signal Processing}, vol. 197, pp. 108553, 2022.

\bibitem{Fesl2022}
Benedikt Fesl, Michael Joham, Sha Hu, Michael Koller, Nurettin Turan, and
  Wolfgang Utschick,
\newblock ``{Channel Estimation based on Gaussian Mixture Models with
  Structured Covariances},''
\newblock in {\em 2022 56th Asilomar Conf. Signals, Syst. Comput.}, 2022, pp.
  533--537.

\bibitem{Kay1993}
Steven~M. Kay,
\newblock {\em {Fundamentals of Statistical Signal Processing: Estimation
  Theory}},
\newblock Prentice-Hall, Inc., Englewood Cliffs, NJ, 1993.

\bibitem{Foucart2013}
Simon Foucart and Holger Rauhut,
\newblock {\em {A Mathematical Introduction to Compressive Sensing}},
\newblock Applied and Numerical Harmonic Analysis. Springer New York, New York,
  NY, 2013.

\bibitem{3GPP2020}
3GPP,
\newblock ``{Spatial channel model for Multiple Input Multiple Output (MIMO)
  simulations (Release 16)},''
\newblock Tech. {R}ep. 25.996 V16.0.0, 3rd Generation Partnership Project
  (3GPP), 2020.

\bibitem{Jaeckel2014}
Stephan Jaeckel, Leszek Raschkowski, Kai Borner, and Lars Thiele,
\newblock ``{QuaDRiGa: A 3-D Multi-Cell Channel Model With Time Evolution for
  Enabling Virtual Field Trials},''
\newblock {\em IEEE Trans. Antennas Propag.}, vol. 62, no. 6, pp. 3242--3256,
  2014.

\bibitem{3GPP2023}
3GPP,
\newblock ``{Physical channels and modulation (Release 17)},''
\newblock Tech. {R}ep. 38.211 V17.5.0, 3rd Generation Partnership Project
  (3GPP), 2023.

\bibitem{Kingma2019}
Diederik~P. Kingma and Max Welling,
\newblock ``{An Introduction to Variational Autoencoders},''
\newblock {\em Found. Trends{\textregistered} Mach. Learn.}, vol. 12, no. 4,
  pp. 307--392, 2019.

\bibitem{Kingma2014}
Diederik~P. Kingma and Max Welling,
\newblock ``{Auto-Encoding Variational Bayes},''
\newblock in {\em Proc. 2nd Int. Conf. Learn. Represent.}, Banff, Canada, 2014.

\bibitem{Gray2005}
Robert~M. Gray,
\newblock ``{Toeplitz and Circulant Matrices: A Review},''
\newblock {\em Found. Trends{\textregistered} Commun. Inf. Theory}, vol. 2, no.
  3, pp. 155--239, 2005.

\bibitem{Bergstra2012}
James Bergstra and Yoshua Bengio,
\newblock ``{Random Search for Hyper-Parameter Optimization},''
\newblock {\em J. Mach. Learn. Res.}, vol. 13, no. 10, pp. 281--305, 2012.

\end{thebibliography}

\end{document}